\documentclass[twocolumn,aps,prl,showpacs,superscriptaddress,nofootinbib]{revtex4}

\usepackage[latin1]{inputenc}
\usepackage{graphicx,epstopdf}
\usepackage{amsmath}
\usepackage{hyperref}
\usepackage{amsfonts}
\usepackage{amssymb}
\usepackage{color}
\usepackage{latexsym}
\usepackage{times,txfonts}

\newcommand{\fig}[1]{{Fig.}}

\begin{document}

\title{Observation of Environment-Induced Double Sudden Transitions in Geometric Quantum Correlations}

\author{F. M. Paula}
\affiliation{Instituto de F\'isica, Universidade Federal Fluminense, Av. Gal. Milton Tavares de Souza s/n, Gragoat\'a, 24210-346, Niter\'oi, RJ, Brazil}

\author{I. A. Silva}
\affiliation{Instituto de F\'isica de S\~ ao Carlos, Universidade de S\~ ao Paulo, Caixa Postal 369, 13560-970 S\~ ao Carlos, S\~ ao Paulo, Brazil}

\author{J. D. Montealegre}
\affiliation{Instituto de F\'isica, Universidade Federal Fluminense, Av. Gal. Milton Tavares de Souza s/n, Gragoat\'a, 24210-346, Niter\'oi, RJ, Brazil}

\author{A. M. Souza}
\affiliation{Centro Brasileiro de Pesquisas F\'isicas, Rua Dr. Xavier Sigaud 150, 22290-180 Rio de Janeiro, Brazil}

\author{E. R. deAzevedo}
\affiliation{Instituto de F\'isica de S\~ ao Carlos, Universidade de S\~ ao Paulo, Caixa Postal 369, 13560-970 S\~ ao Carlos, S\~ ao Paulo, Brazil}

\author{R. S. Sarthour}
\affiliation{Centro Brasileiro de Pesquisas F\'isicas, Rua Dr. Xavier Sigaud 150, 22290-180 Rio de Janeiro, Brazil}

\author{A. Saguia}
\affiliation{Instituto de F\'isica, Universidade Federal Fluminense, Av. Gal. Milton Tavares de Souza s/n, Gragoat\'a, 24210-346, Niter\'oi, RJ, Brazil}

\author{I. S. Oliveira}
\affiliation{Centro Brasileiro de Pesquisas F\'isicas, Rua Dr. Xavier Sigaud 150, 22290-180 Rio de Janeiro, Brazil}

\author{D. O. Soares-Pinto}
\affiliation{Instituto de F\'isica de S\~ ao Carlos, Universidade de S\~ ao Paulo, Caixa Postal 369, 13560-970 S\~ ao Carlos, S\~ ao Paulo, Brazil}

\author{G. Adesso}
\affiliation{School of Mathematical Sciences, The University of Nottingham, University Park, Nottingham NG7 2RD, United Kingdom}

\author{M. S. Sarandy}
\affiliation{Instituto de F\'isica, Universidade Federal Fluminense, Av. Gal. Milton Tavares de Souza s/n, Gragoat\'a, 24210-346, Niter\'oi, RJ, Brazil}

\date{\today}

\begin{abstract}
Correlations in quantum systems exhibit a rich phenomenology under the effect of various sources of noise.
We investigate theoretically and experimentally the dynamics of quantum correlations
and their classical counterparts in two  nuclear magnetic resonance setups, as measured by geometric quantifiers based on trace-norm.
We consider two-qubit systems prepared in Bell diagonal states, and perform the experiments in real decohering environments resulting from
Markovian local noise which preserves the Bell diagonal form of the states. We then report the first
observation of environment-induced double sudden transitions in the geometric quantum correlations, a genuinely nonclassical effect not observable in classical correlations. The evolution of classical correlations in our physical implementation reveals in turn the finite-time relaxation to a pointer basis under nondissipative decoherence, which we characterize geometrically in full analogy with predictions based on entropic measures.

\end{abstract}

\pacs{03.67.Mn, 03.65.Ta, 03.65.Ud, 03.65.Wj}

\maketitle

{\it Introduction.}---Quantum technology  embraces some of the most exciting modern developments in physics, mathematics, engineering and computer science. These were spurred by the discovery that distinctive quantum features, such as superposition and entanglement, can be exploited as practical resources for novel or enhanced sensing, communication and computation \cite{Nielsen-Chuang}.   Mass adoption of quantum technologies, however,  demands a deeper understanding of the ultimate ingredients which enable improvements over classical scenarios, and the ability to sustain those properties against the typically detrimental effects of decoherence. Beyond entanglement, subtler forms of nonclassical correlations such as the quantum discord \cite{Ollivier,Henderson:01} have emerged as key quantum signatures \cite{Modi,Celeri,Sarandy:12} with operational roles e.g.~in quantum metrology \cite{modix,lqu,blind}, entanglement activation \cite{streltsov,Piani:11,sciarrino}, information encoding and distribution \cite{gu,sharing}. Studying the behaviour of quantum correlations under physical sources of noise is then pivotal to their exploitation for real-world implementation of these and related protocols.

In presence of decohering environments, quantum discord does not exhibit the phenomenon of sudden death~\cite{Werlang:09,Wang:10,Ferraro:10},
and is typically more robust against noise than entanglement~\cite{Yu:09} (see also \cite{campbell,fanchininjp}). When evaluated on simple Bell diagonal states of two qubits, the traditional entropic measure of quantum discord \cite{Ollivier} displays peculiar properties in its decay rate,  such as freezing~\cite{Mazzola:10} and sudden changes, i.e.~abrupt transitions at specific evolution times~\cite{Maziero:09} (see \cite{fanchininew} for a critical study on more general states). Moreover, the entropic classical correlation~\cite{Henderson:01} complementary to the quantum discord has been recently employed to characterize the emergence of the pointer basis of an apparatus subject to decoherence~\cite{Cornelio:12}. This characterization shows that the time $\tau_E$ for the emergence of the pointer basis is conceptually distinct from the decoherence half-life  $\tau_D$ (decoherence time). Experimental investigations of both the freezing and the sudden change phenomena have been realized through different physical setups~\cite{Xu:10-1,Xu:10-2,Auccaise:11}, while the emergence of a pointer basis has only been recently observed in an experiment with polarization entangled photon pairs~\cite{Cornelio:12}.

In alternative to an entropic approach, quantum correlations in a bipartite quantum state $\rho$ can also be measured---and typically more simply evaluated---via a geometric scenario, in terms of the distance between $\rho$ and the closest state with zero discord, where states with zero discord are described by density operators which are left undisturbed by a non-selective measurement over one subsystem.  In this context, the trace-norm (Schatten 1-norm) geometric quantum discord (GQD-1) plays a special role \cite{Hu:12,Rana:13,Debarba,Paula,Nakano:12,Ciccarello}, as it defines a rigorous and physically motivated measure which does not increase under local trace-preserving quantum channels for the unmeasured part (as a consequence of the contractivity of the trace-norm \cite{Hu:12,Tufarelli:12,Piani}). The GQD-1 thus does not suffer from an inherent problem which affects instead the Hilbert-Schmidt (Schatten 2-norm) geometric quantum discord originally introduced in~\cite{Dakic}, as well as any Schatten $p$-norm measure of discord for $p\ne 1$. The GQD-1 furthermore inherits the robustness of the entropic quantum discord against noise, being also capable of exhibiting freezing~\cite{Aaronson} and sudden change behaviors~\cite{Montealegre:13}. Moreover, it displays a new effect for Bell diagonal states, which is the possibility of double sudden changes~\cite{Montealegre:13}. The freezing and single sudden change behaviors for GQD-1 have recently been experimentally verified~\cite{Silva:12} through the observation of the negativity of quantumness~\cite{Piani:11}, a quantum correlation measure equivalent to GQD-1 for qubit states~\cite{Paula,Nakano:12}.

In this Letter we report the first experimental realization of the double sudden change behavior via two distinct  Nuclear Magnetic Resonance (NMR) setups, where different Markovian noisy channels are induced by real physical interactions inducing relaxations on partially polarized nuclear spin ensembles. Moreover, we show that these double sudden changes are a genuinely quantum effect, being forbidden in the geometric classical correlations
associated with GQD-1~\cite{Paula:2,Aaronson:2}. Geometric quantum and classical correlations take very simple and experimentally-friendly expressions for Bell diagonal states, directly given by spin-spin correlation functions in different directions. Our study further allows the establishment of a complete equivalence between the entropic and the geometric characterizations of the emergence of the pointer basis in an apparatus subject to decoherence. The relaxation to the pointer basis at a finite emergence time $\tau_E$ is then demonstrated as a by-product in our NMR implementation.

{\it Geometric correlations.}--- Let us begin by considering a two-qubit system $AB$ described by a density operator $\rho$. 
We will focus here in the particular case of two-qubit Bell diagonal states, whose density operator
is of the form $\rho=\frac{1}{4}\left[\mathbb{I}\otimes \mathbb{I}+\sum_{i=1}^{3}c_i\sigma_i\otimes\sigma_i\right]$,
where $\mathbb{I}$ is the identity matrix, $\{\sigma_i\}$ are the Pauli matrices, and $\{c_{i}=\langle\sigma_{i}\otimes\sigma_{i}\rangle\}$ are spin-spin correlation functions. Geometric quantifiers of quantum (namely, GQD-1) and classical correlations between $A$ and $B$ can be defined through the trace distances~\citep{Paula:2}
\begin{eqnarray}
Q_{G}(\rho)&=&\inf_\chi\text{tr}\left|\rho-\chi\right| = \text{tr}\left|\rho-\chi_\rho\right|, \label{QG} \\
C_{G}(\rho)&=&\text{tr}\left|\chi_\rho-\pi_{\rho}\right|, \label{CG}
\end{eqnarray}
where $\chi_\rho$ is a classical-quantum state \cite{notecq} emerging from a projective measurement on
subsystem $A$ that minimizes Eq.~(\ref{QG}), whereas $\pi_{\rho}=\rho_{A}\otimes \rho_{B}$ represents the product of the local marginals of $\rho$. In this case, $\rho_{A}=\rho_{B}=\mathbb{I}/2$ and $\chi_\rho=\left(\mathbb{I}\otimes \mathbb{I}+c_{j}\sigma_{j}\otimes\sigma_{j}\right)/4$, where $j$ is such that $|c_{j}|$
corresponds to the maximum of the set $\{|c_{i}|\}$. Defining $c_{-}$, $c_{0}$, and $c_{+}$ as the \textit{minimum}, \textit{intermediate}, and \textit{maximum} of $\{|c_{i}|\}$, respectively, Eqs.~(\ref{QG}) and (\ref{CG}) reduce to~\citep{Paula:2}
\begin{equation}\label{QCG}
Q_{G}= c_0,\,\,\,\,C_{G}= c_+.
\end{equation}
It is worth mentioning that the classical part in Eq.~(\ref{QCG}) is monotonically related to the expression obtained in Ref.~\cite{Aaronson:2} through an alternative definition for the geometric trace-norm classical correlation, which takes into account an extra optimization to find out the closest product state.

{\it Decoherence.}---We will consider the system-environment interaction through the operator-sum representation formalism~\cite{Nielsen-Chuang}. In this context, we will take the evolution of the quantum state
$\rho$ as described by a trace-preserving quantum channel given by $\varepsilon(\rho) = \sum_{i,j} \left(E_i^{A}\otimes E_j^{B}\right) \rho \left(E_i^{A} \otimes E_j^{B}\right)^\dagger,$
where $\{E_k^{s}\}$ is the set of Kraus operators associated to a decohering process of the qubit $s$,
with the trace-preserving condition reading $\sum_{k} E_{k}^{s\dagger} E_{k}^{s} = \mathbb{I}$. We provide in Table~\ref{t1} the Kraus operators for the two quantum channels considered in this work: {\it Phase Damping} (PD) and {\it Generalized Amplitude Damping} (GAD).

\begin{table}[hbt]
\begin{tabular}{lc}
\hline \hline
Channel & Kraus operators                                         \\ \hline \\[-0.2cm] 
PD   & $E_0^{s} = \sqrt{1-p_{s}/2}\, \mathbb{I}\, ,\  E_1^{s} = \sqrt{p_{s}/2}\, \sigma_3\,.$          
\\  & \\[0.1cm]
&
$E_0^{s}=\sqrt{\gamma_{s}}\left[
\begin{array}{cc}
1 & 0 \\
0 & \sqrt{1-p_{s}} \\
\end{array} \right] ,\
E_2^{s}=\sqrt{1-\gamma_{s}}\left[
\begin{array}{cc}
\sqrt{1-p_{s}} & 0 \\
0 & 1 \\
\end{array} \right] ,$  \\
 GAD  & \\
 & $E_1^{s}=\sqrt{\gamma_{s}}\left[
\begin{array}{cc}
0 & \sqrt{p_{s}} \\
0 & 0 \\
\end{array} \right] ,\
E_3^{s}=\sqrt{1-\gamma_{s}}\left[
\begin{array}{cc}
0 & 0 \\
\sqrt{p_{s}} & 0 \\
\end{array} \right] .$  \\[.4cm] \hline \hline
\end{tabular}
\caption[table1]{Kraus operators for the PD and GAD quantum channels, where the parameters
$p_{s}$ and $\gamma_{s}$ represent decoherence probabilities.}
\label{t1}
\end{table}

The PD decoherence process preserves the Bell diagonal form of the
density operator $\rho$. For the case of GAD, the Bell diagonal form is kept for $\gamma_{s}=1/2$. In this situation, as $p_{s}=1-\text{exp}\left[-t/T_s\right]$ is a function time $t$, being $T_s$ the relaxation time of the
qubit $s$, we can write
\begin{equation}\label{rho-t}
\varepsilon(\rho)=\rho(t)=\begin{array}{c}\frac{1}{4}\left[\mathbb{I}\otimes \mathbb{I}+\sum_{i=1}^{3}c_i(t)\sigma_i\otimes\sigma_i\right]\end{array},
\end{equation}
where the time-dependent correlation  function $c_i(t)$ is given in Table~\ref{t2} in terms of the initial value $c_i(0)= c_i$ and of the \textit{decoherence time} $\tau_D=T_AT_B/(T_A+T_B)$.

Since $\rho(t)$ preserves the Bell diagonal form, we can directly obtain
the dynamics of the geometric quantum and classical correlations from Table~\ref{t2} by using the respective expressions
$Q_{G}(t)= c_0(t)$, $C_{G}(t)= c_+(t)$.
Thus, we can get general patterns of the geometric correlations as functions of the parameters $\{c_{j}\equiv c_{j}(0)\}_{j=-,0,+}$ and $\tau_D$. Depending on the values assumed by these parameters, $Q(t)$ and $C(t)$ may present non-analyticities caused by crossings among $|c_1(t)|$, $|c_2(t)|$, and $|c_3(t)|$, which give rise to  the occurrence of double sudden change and emergence of the pointer basis phenomena.

\begin{table}[b]
\begin{tabular}{lccc}
\hline \hline
Channel                   & $\qquad \ \  c_1(t) \ \  \qquad$      & $\qquad \  \ c_2(t)\  \ \qquad$     & $\qquad\ \  c_3(t)\ \  \qquad$      \\ \hline 
PD                 &  $c_1 \text{exp}\left[-t/\tau_D\right]$            & $c_2 \text{exp}\left[-t/\tau_D\right]$    & $c_3$     \\
GAD                &  $c_1 \text{exp}\left[-t/2\tau_D\right]$ & $c_2 \text{exp}\left[-t/2\tau_D\right]$ & $c_3 \text{exp}\left[-t/\tau_D\right]$ \\
\hline \hline
\end{tabular}
\caption[table2]{Correlation function $c_i(t)$ for the PD and GAD quantum channels in terms of $c_i$ and $\tau_D$. For GAD, we fixed $\gamma_s=1/2$.}
\label{t2}
\end{table}

{\it Double sudden change.}---For a two-qubit Bell diagonal state evolving under
Markovian local noise preserving their Bell diagonal form as described by the PD or GAD channels, the quantum geometric correlation (GQD-1) is able to exhibit at most two sudden changes as a function of time. Indeed, it can be observed in Table~\ref{t2} that, for both PD and GAD channels, $|c_1(t)|$ and $|c_2(t)|$ display the same decay rate, which means that they do not cross as functions of $t$. Therefore, only the crossings $|c_3(t)|=|c_1(t)|$ and $|c_3(t)|=|c_2(t)|$ are allowed, implying at most two non-analyticities in the intermediate value $c_0(t)=Q_G(t)$. The analytical expressions for the critical points $t=t_1^{*}$ and $t=t_2^{*}$ ($t_2^{*}> t_1^{*}$) associated with this double sudden change behavior are indicated in Table~\ref{t3}. These results generalize the expressions obtained in Ref.~\cite{Montealegre:13} for the particular case $T_A=T_B$.
\begin{table}[hbt]
\begin{tabular}{lccc}
\hline \hline
Channel                   & $\qquad \quad t_1^{*}\quad \qquad$     & $\qquad \quad t_2^{*}\quad \qquad $    &$\qquad  \mbox{Conditions} \qquad$     \\ \hline
& & & \\[-0.3cm]
PD                 &  $\tau_D\text{ln}\left[c_0/|c_3|\right]$    & $\tau_D\text{ln}\left[c_+/|c_3|\right]$ & $|c_3|=c_-$ \\ & & &  $c_+ \neq c_0 \neq c_- \neq 0$    \\[.2cm]
GAD                &   $2\tau_D\text{ln}\left[|c_3|/c_0\right]$  &   $2\tau_D\text{ln}\left[|c_3|/c_-\right]$ &  $|c_3|=c_+$ \\ & & &   $c_+ \neq c_0 \neq c_- \neq 0$ \\[0.1cm] \hline \hline
\end{tabular}
\caption[table3]{Critical points $t_1^{*}$ and $t_2^{*}$ in terms of the parameters $c_{-}$, $c_0$, $c_+$, and $\tau_D$.The conditions provided are necessary and sufficient for the occurrence of double sudden transitions.}
\label{t3}
\end{table}

On the other hand, the crossings $|c_3(t)|=|c_1(t)|$ and $|c_3(t)|=|c_2(t)|$ allows for at most a single non-analyticity in the geometric classical correlations $C_G(t)=c_+(t)$. This sudden change occurs at times $t_{2}^{*}$ and $t_{1}^{*}$ for the PD and GAD channels under the necessary and sufficient conditions $c_+ > |c_3|\neq 0$ and $|c_3|>c_0\neq 0$, respectively.
This implies that the double sudden change is, for Bell diagonal states, a genuinely quantum effect, which is unattainable
for $C_G$. Since the original entropic classical correlation $C_E$  \cite{Henderson:01} is monotonically related with $C_G$ within the whole class of Bell diagonal states~\cite{Paula:2}, its dynamics $C_E(t)$ also presents at most one single sudden change. Moreover, as the original entropic quantum discord \cite{Ollivier} is given by $Q_E=T_E-C_E$, being the entropic total correlation $T_E$ a smooth function, we conclude that $Q_E(t)$ is unable to display double sudden changes.

{\it Emergence of the pointer basis.}---A quantum apparatus ${\cal A}$ measuring a system ${\cal S}$ suffers decoherence through contact with the environment, collapsing into a possible set of classical states known as {\it pointer basis}~\cite{Zurek:81}. The emergence of these pointer states is attributed to the instant of time $\tau_E$ at which the entropic classical correlation $C_E$ between ${\cal A}$ and ${\cal S}$ becomes constant~\cite{Cornelio:12}. In other words, $\tau_E$ is the necessary time for the information about ${\cal S}$ to be accessible to a classical observer. In particular, if the system ${\cal AS}$ under decoherence is described by Eq.~(\ref{rho-t}), the geometric classical correlation $C_G$ can be used to characterize $\tau_E$, since $C_G$ and $C_E$ are monotonic between themselves for Bell diagonal states. In fact, from Table~\ref{t2} it is easily seen that $C_G(t)=c_0(t)$ becomes constant in finite time only for the PD channel, at the sudden change point
\begin{equation}
\tau_{E}=\tau_D\,\text{ln}[{c_+}/{|c_3|}] \qquad (c_+ > |c_3|\neq 0),
\label{taue}
\end{equation}
which is in complete agreement with $\tau_{E}$ found in Ref.~\cite{Cornelio:12} by using $C_E$. Depending on the ratio $c_+/|c_3|$, the pointer basis can emerge at time $\tau_E$ smaller or larger than the decoherence time $\tau_D$.
On the other hand, the GAD channel does not allow for the emergence of a pointer basis at finite time.

\begin{figure}[ht!]
\includegraphics[scale=0.35]{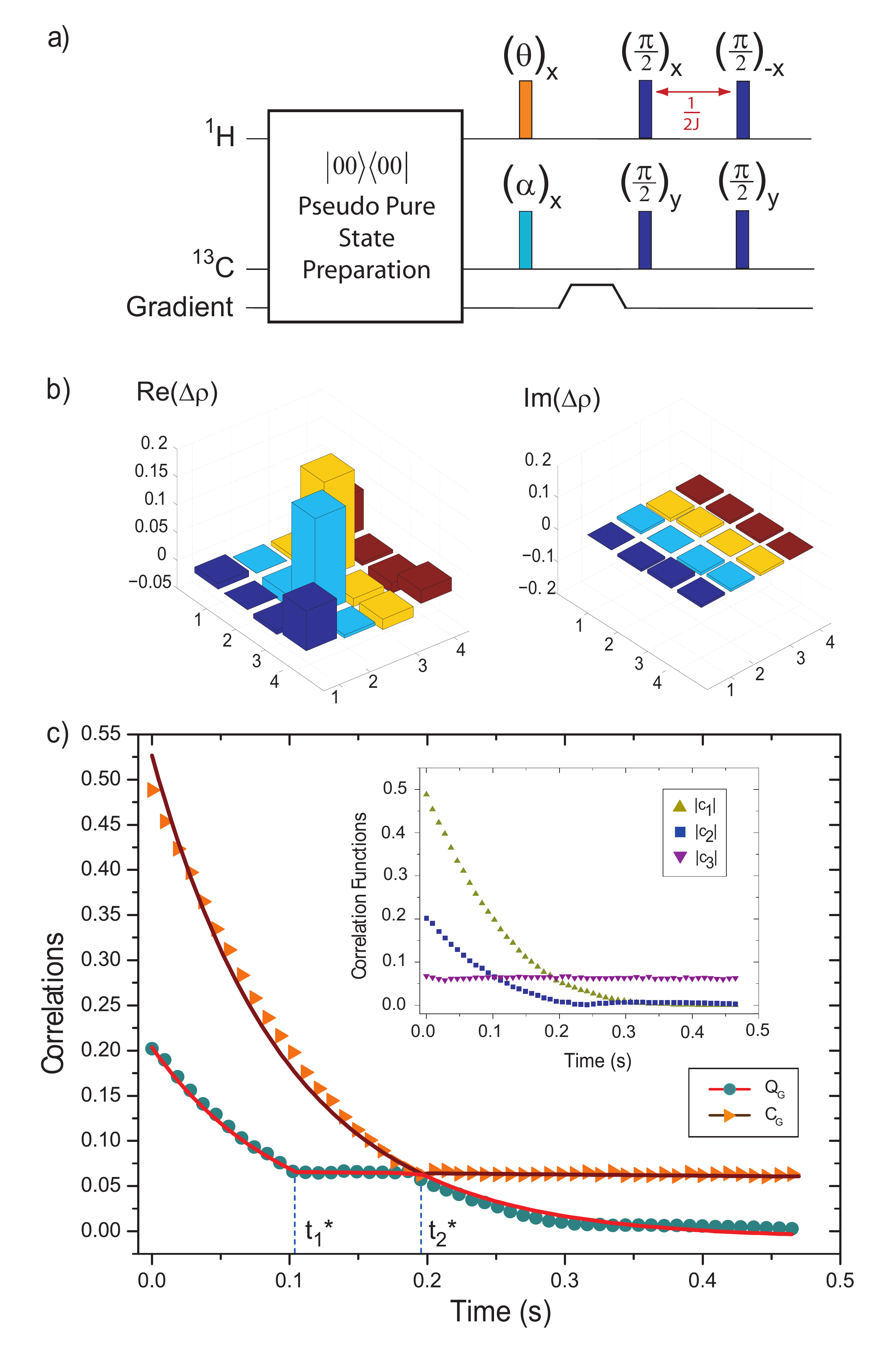} \caption{(Color online) (a) Schematic representation of the pulse sequence employed to obtain a deviation matrix in the form of a Bell diagonal state. (b) Experimentally reconstructed block diagrams for real and imaginary parts of the deviation matrix related to the Bell diagonal initial state with $c_1 = 0.49$, $c_2 = 0.20$ and $c_3 = 0.067$. The curves in (c) denote the time evolutions of quantum ($Q_G$, bullet) and classical ($C_G$, triangle) correlations, respectively. The dots represents the experimental results and the solid lines are the theoretical predictions. In the inset we detail the time evolutions of $|c_1|$ (yellow upward triangles), $|c_2|$ (blue squares), and $|c_3|$ (purple downward triangles) experimentally obtained for the PD decoherence process.}\label{figPD}
\end{figure}

{\it Experimental results.}--- We performed two distinct experimental implementations based on NMR setups, each of them realizing different physical quantum channels. Our aim was to explore separately either the PD or GAD channels, since they are inequivalent concerning the emergence of the pointer basis. The state of a NMR two-qubit system in the high temperature approximation is given by $\rho=\frac{1}{4}\mathbb{I}_{4}+\epsilon\Delta\rho$, where $\epsilon=\hbar\omega_L/4k_BT\sim10^{-5}$ is the ratio between the magnetic and thermal energies at room temperature, $\omega_L$ is the Larmor frequency, $k_B$ is the Boltzmann constant and $T$ the temperature \cite{abra,ivan}. All measurements and transformations affect only the deviation matrix $\Delta\rho$, which contains the available information about the system state. The unitary operations over $\Delta\rho$ are implemented by radiofrequency pulses and evolutions under spin interactions, with an excellent control of rotation angle and phase. Additionally, NMR offers very reliable quantum state tomography, which can be used to obtain a full characterization of $\Delta\rho$ \cite{long2001,leskowitz2004,teles2007}. It is worth remarking that, since in NMR experiments only the deviation matrix is detected, the calculations of the correlation matrix elements $\{c_i\}$ are done in units of $\epsilon$.

The first set of NMR experiments were performed on a liquid state carbon-13 enriched chloroform sample (CHCl$_{3}$) at room temperature, with the two qubits being encoded in the $^{1}$H and $^{13}$C spin--$\frac12$ nuclei, and employing a Varian $500$ MHz spectrometer. The relevant relaxation times were estimated as $T^C_1 \approx 12.46$ s; $T^{C*}_2 \approx 0.15$ s for 
$^{13}$C, and $T^H_1 \approx 7.53$ s; $T^{H*}_2\approx 0.27$ s, for $^{1}$H. Since $T_1 \gg T^*_2$ for both qubits and the system decoherence was evaluated within a maximum evolution time of approximately $0.5$ s, the relaxation mechanism is described effectively by a PD channel with $T_A = T^{H*}_2$ and $T_B=T^{C*}_2$. A Bell diagonal state, with the coefficients $c_1 = 0.49$, $c_2 = 0.20$, and $c_3 = 0.067$,  was obtained applying the pulse sequence illustrated in  \fig{}\ref{figPD}a, where the pseudo pure state $|00\rangle\langle 00|$ was prepared as described in Refs.~\cite{Knill,Nielsen-Chuang}. The
 resulting deviation matrix was of the form  $\Delta\rho=c_1\sigma_x\otimes\sigma_x+c_2\sigma_y\otimes\sigma_y+c_3\sigma_z\otimes\sigma_z$ where $c_1=-2\cos\theta,c_2=-2\cos\alpha,c_3=-2\cos\theta\cos\alpha$, with tunable $\alpha$ and $\theta$. After the initial state preparation, the system was left to evolve for a time $\tau_i$ and quantum state tomography \cite{leskowitz2004} was performed to reconstruct the evolved deviation matrix. The process was repeated for different values of $\tau_i$ in order to monitor the state relaxation. \fig{} \ref{figPD} shows the time dependence of the quantum and classical geometric correlations as well as the individual correlation functions, calculated from the measured deviation matrices. The continuous lines correspond to theoretical predictions calculated according to the expressions in Table~\ref{t2}. Note that $Q_G$ clearly exhibits a double sudden transition at critical points $t_1^*=(0.107 \pm 0.006)$s and $t_2^*=(0.198 \pm 0.006)$s, which are compatible with the theoretical values $t_1^* = 0.105$s and $t_2^* = 0.192$s, as predicted by Table~\ref{t3}. The emergence of the pointer basis in $C_G$ is also evident at $\tau_E = t_2^*$.

\begin{figure}[ht!]
\includegraphics[scale=0.35]{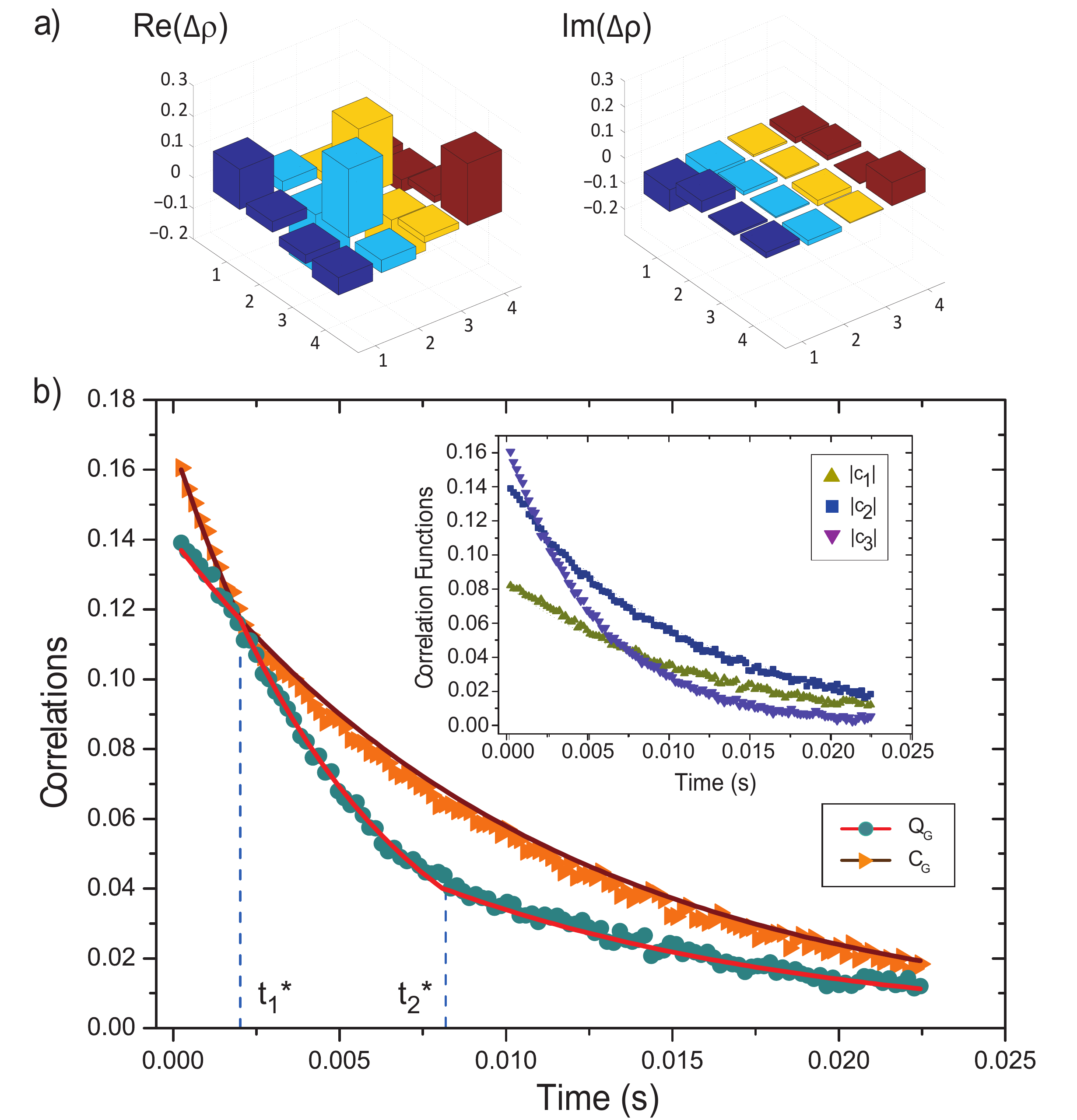} \caption{\label{figGAD} (Color online) (a) Experimentally reconstructed block diagrams for real and imaginary parts of the deviation matrix related to the Bell diagonal initial state with $c_1 = 0.08$, $c_2 = 0.14$ and $c_3 = 0.16$. The curves in (b) denote the time evolutions of quantum ($Q_G$, bullet) and classical ($C_G$, triangle) correlations, respectively. The dots represents the experimental results and the solid lines are the theoretical predictions. In the inset we detail the time evolutions of $|c_1|$ (yellow upward triangles), $|c_2|$ (blue squares), and $|c_3|$ (purple downward triangles) experimentally obtained for the GAD decoherence process.}
\end{figure}

The implementation of a system in which the relaxation is fully described by a GAD channel was achieved using a spin--$\frac32$ NMR quadrupolar system. A spin--$\frac32$  in the presence of a strong static magnetic field is described by four energy levels, which can be indexed as $|00\rangle$,$|01\rangle$, $|10\rangle$, $|11\rangle$ to represent a two-qubit system \cite{kitrin,kumar,teles2007,soarespintoquad}. The energy separation between the levels as well as the system relaxation are dictated by the interaction between the nuclear quadrupole moments with electric field gradients produced by the charge distribution in their surroundings \cite{abra}. The relaxation for this system is described by two-channels: a GAD channel as in Table~\ref{t1} and a global PD channel, acting simultaneously on both logical qubits, with  Kraus operators given in Ref.~\cite{souzaam}. The global PD channel does not act on the cross-diagonal terms of the deviation matrix, which means that for Bell diagonal states the decoherence is completely dictated by the GAD channel \cite{souzaam}.
The experiments on the spin--$\frac32$ system were performed with sodium nuclei in a liquid crystal sample at room temperature using a Varian $400$ MHz spectrometer.  The relevant relaxation times were estimated as $T_A  = T_B \approx 0.012$ s and the initial Bell diagonal state ($c_1 = 0.08$, $c_2 = 0.14$, and $c_3 = 0.16$, \fig{}\ref{figGAD}a) was prepared using a set of numerically optimized radiofrequency pulses obtained by the strong modulated pulse technique \cite{fortunato}.
\fig{} \ref{figGAD} shows the time dependence of the quantum and classical correlations as well as the correlation functions, as calculated from the measured deviation matrices following the same procedure described before.
Note also here that $Q_G$ exhibits a double sudden transition, corresponding to the crossings $|c_2(t)|=|c_3(t)|$ and $|c_1(t)|=|c_3(t)|$, at  critical points $t_1^*=(0.0020\pm 0.0005)$s and $t_2^*=(0.0081\pm 0.0005)$s. According to Table~\ref{t3}, the theoretical critical points are $t_1^* = 0.0016$s and $t_2^* = 0.0083$s. As expected, the presence of a single sudden change in the $C_G$ curve is also observed, but there is no emergence of pointer basis at finite time in this case.

{\it Conclusions.}---We performed a detailed analysis of {\it bona fide} geometric measures of quantum and classical correlations in paradigmatic two-qubit Bell diagonal states realized experimentally by room temperature NMR setups, in the presence of physical sources of phase damping and generalized amplitude damping noise. We observed recently predicted distinctive features of quantum correlations such as double sudden changes in their dynamics \cite{Montealegre:13}, and saturation of classical correlations denoting a relaxation to a pointer basis \cite{Cornelio:12}.
Future investigations will be devoted to assess whether the finite-time emergence of the pointer basis, here established for geometric correlations complementing the case of entropic ones \cite{Cornelio:12}, can be regarded as a universal feature common to all valid quantifiers of classical correlations, along the lines of \cite{Aaronson}.

{\it Acknowledgments.}--- This work is supported by the Brazilian agencies CNPq, CAPES, FAPERJ, FAPESP, the Brazilian National Institute for Science and Technology of Quantum Information (INCT-IQ), and the University of Nottingham.

\end{document}